\documentstyle[aps,prl,floats,epsf]{revtex}
\epsfclipon
\newcommand{\smeq}{\! \! = \!}
\newcommand{\smneq}{\! \! \neq \!}

\begin{document}
\twocolumn[{
\draft
\widetext

\author{Harold U. Baranger$^{1}$ and Pier A. Mello$^{2}$}

\address{$^{1}$ Bell Laboratories--- Lucent Technologies,
700 Mountain Ave. 1D-230, Murray Hill NJ 07974}
\address{$^{2}$ Instituto de F\'{\i}sica, Universidad Nacional Aut\'{o}noma
de M\'{e}xico, 01000 M\'{e}xico D.F., Mexico}

\title{Reflection Symmetric Ballistic Microstructures: 
Quantum Transport Properties}

\date{Submitted to Phys. Rev. Lett. 22 July 1996}
\maketitle

\mediumtext
\begin{abstract}
We show that reflection symmetry has a strong influence on quantum transport
properties. Using a random $S$-matrix theory approach, we derive the
weak-localization correction, the magnitude of the conductance fluctuations,
and the distribution of the conductance for three classes of reflection
symmetry relevant for experimental ballistic microstructures.  The
$S$-matrix ensembles used fall within the general classification scheme
introduced by Dyson, but because the conductance couples blocks of the
$S$-matrix of different parity, the resulting conductance properties are
highly non-trivial.

\end{abstract}
\pacs{PACS numbers: 72.20.My, 05.45.+b, 72.15.Gd}
}]

\narrowtext

The effect of symmetry on the quantum corrections to the conductance is one
of the main themes of mesoscopic physics. Perhaps most familiar is the
suppression of the average quantum conductance at zero magnetic field
because of time-reversal symmetry, an effect known as weak-localization.
Previous work has concentrated on the effects of time-reversal and
spin-rotational invariance on transport through disordered conductors
\cite{RevMes}.

The advent of clean microstructures, in which transport is largely ballistic
\cite{RevMes,Marcus,Chang94}, opens the possibility of studying different
spatial reflection symmetries. In these structures the electrons move in a
potential largely determined by lithography, and structures of definite
reflection symmetry, as well as of adjustable symmetry, are possible.
Alternatively, reflection symmetric scattering systems could be probed in
microwave cavities \cite{microwave}.  Our goal here is to determine how such
reflection symmetries affect the interference contribution to transport.

Because conductance is related to scattering from the system, the symmetry 
classes for quantum transport are closely related to those for the scattering
matrix $S$. It has been shown \cite{uzi,hans,prl94,jlp}
that ensembles in which $S$ is distributed with an
``equal-a-priori probability'' across the available matrix space provide a
good description of the statistical properties of quantum transport when
(1) the classical dynamics is chaotic, 
(2) direct processes through the system are ruled out, and 
(3) there are no spatial symmetries. 
The symmetry classes for such $S$-matrices were
introduced by Dyson \cite{dyson,mehta-porter}.

In Dyson's scheme there are three basic symmetry classes.  In the absence of
any symmetries, the only restriction on $S$ is unitarity due to flux
conservation: this is the {\it unitary} case, denoted $\beta \smeq 2$.  In
the {\it orthogonal} case ($\beta \smeq 1$), $S$ is symmetric because one
has either time-reversal symmetry and integral spin or time-reversal
symmetry, half-integral spin and rotational symmetry.  In the 
{\it symplectic} case ($\beta \smeq 4$), $S$ is self-dual because of
time-reversal symmetry with half-integral spin and no rotational symmetry.
The intuitive idea of ``equal a priori probability'' is expressed
mathematically by the measure on the matrix space which is invariant under
the symmetry operations for the class in question. This notion of 
{\it invariant measure} gives rise to the {\it circular } orthogonal, unitary 
and symplectic {\it ensembles (COE, CUE, CSE)}.

In the presence of additional symmetries, for fixed values for all quantum
numbers of the full symmetry group, the invariant ensemble is one of the
three circular ensembles \cite{dyson}. Thus for reflection symmetries, $S$
is block diagonal in a basis of definite parity with respect to reflection,
with a circular ensemble in each block. This is the natural representation 
for the eigenvalues of $S$, and the statistics of eigenvalues for such
independent superpositions have been studied previously
\cite{dyson,mehta-porter}.  However, the conductance of a system is 
{\it not} an eigenvalue property: it depends on the properties of the
scattering {\it wave-functions}, and those appropriate for the conductance 
are not necessarily of definite reflection parity. Hence the conductance 
may couple the different
parity-diagonal blocks of $S$, and the resulting quantum transport
properties are a non-trivial generalization of the circular ensemble
results.

In this paper we investigate the effects of three classes of reflection
symmetry on the quantum scattering properties of classically chaotic systems
\cite{hastings}.  We consider two-dimensional systems with spinless
particles and rule out any direct processes. The three classes are 
(1)~{\it Up-down (UD)---} reflection through an axis parallel to the current, 
(2)~{\it Left-right (LR)---} reflection through an axis perpendicular to the
current, and (3)~{\it Four-fold (4F)---} the combination of up-down and
left-right. The resulting statistical distribution of the transmission $T$
through the system is calculated. The predictions compare favorably with
numerical calculations in which the Schr\"odinger equation is solved for a
number of structures.

{\it Structure of the $S$-matrix---}
Consider (spinless) single-electron scattering by a ballistic
quantum dot connected to the outside by two leads, each supporting $N$ 
propagating modes. The $2N$-dimensional $S$ matrix, which relates the 
incoming amplitudes from the left and right $\lbrace a^{L,R} \rbrace$ 
to the outgoing ones $\lbrace b^{L,R} \rbrace$, is
\begin{equation}
\label{defnS}
\left[ 
\begin{array}{c}
b^L \\ 
b^R 
\end{array}
\right] =S\left[ 
\begin{array}{c}
a^L \\ 
a^R 
\end{array}
\right] 
\; ; \;  
S=\left[ 
\begin{array}{cc}
r & t^{\prime } \\ 
t & r^{\prime } 
\end{array}
\right] , 
\end{equation} 
where $r$, $r^{\prime }$ are the $N\times N$ reflection matrices for
incidence from each lead and $t$, $t^{\prime }$ the corresponding
transmission matrices. The conductance is given in terms of the
transmission coefficient $T\smeq  {\rm tr} [tt^{\dagger }]$ by 
$G\smeq (e^2/h)T$  \cite{RevMes}.

The reflection symmetries indicated above impose restrictions on $S$: 
the resulting structure, both in the absence and in the presence of a 
magnetic field $B$, is presented in Table I, which we now discuss. 

\begin{table}[tbp]
\caption{The structure of the $S$ matrix for up-down (UD), left-right (LR),
and four-fold (4F) reflection symmetry.}
\begin{tabular}{ccc} 
       &${\bf B=0}$ & ${\bf B\smneq 0}$  \\ \tableline
 \\
 {\bf UD}  
            & $\left[ \begin{array}{cc} 
               \begin{array}{cc} r_{e} & t_{e}^T \\ t_{e} & r_{e}^{^{\prime
}}\end{array}
               & 0 \\ 0 & 
                \begin{array}{cc} r_{o} & t_{o}^T \\ t_{o} & r_{o}^{^{\prime
}}\end{array}
                \end{array} \right] $ 
            & $\left[ \begin{array}{cc} r & t^T \\
t & r^{\prime }
                                  \end{array} \right] $ \\
&     $ r_{e,o }=r_{e,o }^T,\quad r_{e,o }^{^{\prime }}=
                                  r_{e,o}^{^{\prime }T} $
&
                    $ r=r^T,\quad r^{\prime }=r^{{\prime }T} $ \\ 
 \\
 {\bf LR} 
                 & $\left[ \begin{array}{cc} r & t \\ t & r\end{array}
\right] $  
                          & $\left[ \begin{array}{cc} r & t^{{\prime }} \\ t
& r^T \end{array} \right] $  \\
                 & $r=r^T,\quad t=t^T$ 
                                   & $t=t^T,\quad t^{\prime }=t^{{\prime
}T}$ \\ 
 \\
{\bf 4F}   
                      & $\left[ \begin{array}{cc} 
                       \begin{array}{cc} r_{e} & t_{e} \\ t_{e} &
r_{e}\end{array}
                       & 0 \\ 0 & 
                        \begin{array}{cc} r_{o} & t_{o} \\ t_{o} &
r_{o}\end{array} 
                        \end{array} \right] $  
                                     & $\left[ \begin{array}{cc} r & t \\ t
& r\end{array}\right] $ \\
                       & $r_{e,o}=r_{e,o}^T,\quad t_{e,o}=t_{e,o}^T $ 
                                      & $r=r^T,\quad t=t^T$     \\ 
\end{tabular}
\end{table}

First, for $B\smeq 0$, the wave functions in the leads are 
$\exp (ik_n x)\chi _n (y)$ where the channel wave functions $\chi _n (y)$
vanish at the walls of the leads.  For UD symmetry, $S$ contains two
disconnected symmetric $S$ matrices, for the even ($e$) and odd ($o$) parity
channels, respectively.  For LR, changing $x\rightarrow -x$ in the wave
function gives a solution for the same energy: consistency with Eq.
(\ref{defnS}) and $S\smeq S^T$ (from time-reversal invariance) gives the
structure shown.  By transforming to states of definite parity, $S$ in the
LR case becomes block diagonal with the matrices $S_{\pm } \equiv r\pm t$ on
the diagonal; hence $S$ is fundamentally a superposition of two symmetric
$S$ matrices as for UD.  A similar procedure is used to analyze 4F symmetry.

Second, for a uniform $B$ in the $z$-direction,
the Schr\"odinger equation in the Landau gauge is
\begin{equation}
\label{Schr1}
\left[ \frac 1{2m}\left( p_x-\frac ecBy\right) ^2+ 
\frac{p_y^2}{2m}+V\right] \psi \\ =E\psi \; .
\end{equation}
General consequences of the spatial symmetry of $V(x,y)$ now follow. 
For UD symmetry, $V(x,y)\smeq V(x,-y)$, so that $\psi ^{*}(x,-y)$ satisfies 
Eq. (\ref{Schr1}). We thus have the antiunitary
symmetry $R_x \vartheta $, where $R_x$ is the reflection operator
with respect to the $x$ axis and $\vartheta$ is the time-reversal operator.
For LR, $V(x,y)\smeq V(-x,y)$, $\psi ^{*}(-x,y)$ satisfies 
Eq. (\ref{Schr1}), and we have the antiunitary symmetry $R_y\vartheta $, 
with $R_y$ the reflection operator with respect to the $y$ axis. 
For 4F, both $R_x\vartheta $ and $R_y\vartheta $ are relevant.
We now consider a scattering problem in which the basic wave functions 
in the leads are $\exp (ik_n x)\chi _{nk}(y)$. 
For UD symmetry, application of $R_x \vartheta$ leads to  $S\smeq S^T$
\cite{RobBer}.
For LR, $R_y \vartheta$  gives $( \Sigma _xS)\smeq (\Sigma _xS)^T$, where 
$ \Sigma _x \smeq \left[ 
\begin{array}{cc}
0_N & 1_N \\ 
1_N & 0_N 
\end{array}
\right] $ and $0_N$ and $1_N$ are the $N\times N$ zero and unit matrices,
respectively. 
Finally, for 4F one finds $S\smeq S^T$ and $(\Sigma _xS) \smeq (\Sigma _xS)^T$,
so that $S$ has the same structure as for LR at $B\smeq 0$.
These results lead to the structure of $S$ shown in Table I.

{\it Invariant measure---}
The invariant measure is most easily written down in a basis of definite
parity with respect to all symmetry operators. In this case, it is simply a
product of circular ensemble measures $d\mu^{(\beta )} (S)$ for each block
of $S$; results are presented in Table II. 
Thus for $B \smeq 0$, the invariant measure for both UD and LR symmetry
is a product of two COE. For $B \smneq 0$, because of the single antiunitary
symmetry present for both UD and LR, the ensemble is a single COE. For
LR note that {\it the roles of $r$ and $t$ are reversed} from the usual case.

\begin{table}[t] 
\caption{The invariant measure for the $S$ matrix for three 
reflection symmetry classes. $d\mu^{(1)}$ is the COE measure.}
  \begin{tabular}{ccc}
                   &  ${\bf B=0}$  &  ${\bf B\neq 0}$ \\ 
\tableline
   {\bf UD}         &  $d\mu ^{(1)}(S_e)d\mu ^{(1)}(S_o)$
                              &  $d\mu ^{(1)}(S)$  \\ 
   {\bf LR}          &$d\widehat{\mu }^{(1)}(S) \equiv 
                            d\mu ^{(1)}(S_{+})d\mu ^{(1)}(S_{-})$
                                    &  $d\mu ^{(1)}(S)$  \\ 
   {\bf 4F}      & $d\widehat{\mu }^{(1)}(S_e)d\widehat{\mu }^{(1)}(S_o)$      
                                &  $d\widehat{\mu }^{(1)}(S)$ \\ 
  \end{tabular}
\end{table}

{\it Statistical properties of $T$---}
Once the invariant measure is known, the statistical properties of the
transmission are found by integrating over the measure, for instance
$\langle T \rangle \smeq  \int T d\mu (S)$. Such averages can be calculated
explicitly using known properties of unitary and orthogonal matrices
\cite{prl94,jlp,jpa}. In Tables III, IV and V we present results for the
weak-localization correction (WLC) $\left\langle T\right\rangle \! -\! N/2$, 
the variance of $T$, and the probability density $w(T)$. 
We now discuss these results.

\setcounter{topnumber}{3}
\begin{table}[t]
\caption{The weak-localization correction 
$\left\langle T\right\rangle \! - \! N/2$  for three classes of 
reflection symmetry. Note that this quantity is {\it positive} for LR symmetry 
in a magnetic field while the magnetoconductance is zero for 4F symmetry.}
  \begin{tabular}{ccc} 
                   &  ${\bf B=0}$  &  ${\bf B\neq 0}$ \\ 
\tableline
     ${\bf UD}$ & $-\sum_{i=e,o} N_i/(4N_i+2)$   &  $-N/(4N+2)$ \\ 
     ${\bf LR}$  &   $0$   &    $N/(4N+2)$ \\ 
     ${\bf 4F}$      & $0$   &   $0$                       \\ 
            \end{tabular}
\end{table}

\begin{table}[t]
\caption{var $T$ for three classes of reflection symmetry. 
Greater symmetry produces larger fluctuations.}
  \begin{tabular}{ccc} 
                   &  ${\bf B=0}$  &  ${\bf B\neq 0}$   \\ 
\tableline
   ${\bf UD}$         &
$\sum_{i=e,o}\frac{N_i(N_i+1)^2}{(2N_i+1)^2(2N_i+3)} $           
                                                      &
$\frac{N(N+1)^2}{(2N+1)^2(2N+3)}$   \\ 
   ${\bf LR} $         &  $ N/(4N+4)$            
                                                       &
$\frac{N(N+1)^2}{(2N+1)^2(2N+3)}$ \\ 
   ${\bf 4F}$      &   $\sum_{i=e,o} N_i/(4N_i+4)$            
                                                       & $N/(4N+4)$ \\ 
  \end{tabular}
 \end{table}

\begin{table}[b]
  \begin{tabular}{ccc} 
                   &  ${\bf B=0}$  &  ${\bf B \neq 0}$ \\ \tableline
 \\
  ${\bf UD}$ &   $\begin{array}{c} 1/\sqrt{4T}   \\ 
                        \left\{\begin{array}{c} \pi /4 \; ,\\ 
                  \frac 12\sin ^{-1}\frac{2-T}T    \end{array} \right.
                                    \end{array}$         
                      &      $\begin{array}{c} 1/\sqrt{4T}   \\ 
                        \left\{\begin{array}{c} 3T/2 \\ 
                               \frac 32 (T-2\sqrt{T-1})\end{array} \right.
                                   \end{array} $         
                                              \\ 
 \\
 ${\bf LR}$ &   $\begin{array}{c} 1/\sqrt{\pi^2T(1-T)}   \\ 
                        \frac{1}{\pi} \ln \frac{1+\sqrt{T(2-T)}}{|1-T|}
\end{array} $
                  &      $\begin{array}{c} 1/\sqrt{4(1-T)}    \\ 
                        \left\{\begin{array}{c} \frac 32 (2-T-2\sqrt{1-T}) \\
                                      \frac {3}{2} (2-T)  \end{array}
\right.              
                                       \end{array}$
                                           \\ 
 \\
 ${\bf 4F}$   &   $\begin{array}{c} 1/\sqrt{\pi^2T(1-T)}   \\ 
                        \frac{2}{\pi ^2} K(\sqrt{T(2-T)})  \end{array} $
                         &  $\begin{array}{c} 1/\sqrt{\pi^2T(1-T)}   \\ 
                        \frac{1}{\pi} \ln \frac{1+\sqrt{T(2-T)}}{|1-T|}
\end{array} $
                                    \\ 
  \end{tabular}
\caption{Probability distribution $w(T)$ for three classes of reflection 
symmetry. $N \smeq 1$ ($2$) in the upper (lower) half of each entry.
For $N \smeq 2$ the braces contain results for $0 \!<\! T \!<\! 1$ on the 
upper line and for $1 \!<\! T \!<\! 2$ on the lower one. ($K$ is the
complete elliptic integral.)}
\end{table}

For UD symmetry, the independent even and odd channels at $B \smeq 0$
imply that the average and variance of $T$ are simply the sum of the
COE results for each parity class while the distribution is the 
convolution. For $B \smneq 0$, one obtains the usual COE results.

For LR symmetry, at $B\smeq 0$ one has both coherent back-scattering and
coherent {\it forward}-scattering ($t \smeq t^T$), so that the WLC is
identically zero. When $B \smneq 0$, the coherent back-scattering is
destroyed but the coherent forward-scattering remains because of the 
$R_y \vartheta$ symmetry. Thus the average transmission is {\it larger} than 
the classical value $N/2$, an unusual result for spinless particles. In the
probability density $w(T)$ for $N\smeq 1$, the coherent forward scattering
produces a square-root singularity at $T\smeq 1$; thus for $B\smeq 0$,
$w(T)$ has singularities at {\it both} $T\smeq 0$ and $T\smeq 1$.  For
$N\smeq 2$ and $B\smeq 0$, note the logarithmic singularity in $w(T)$ at
$T\smeq 1$.

For 4F symmetry, there is sufficient symmetry so that both coherent
back-scattering and coherent forward-scattering exist at $B \smneq 0$. Thus
the WLC is identically zero at both $B\smeq 0$ and $B \smneq 0$, and so 
{\it the average magnetoconductance is zero}.  Finally, notice the large 
variance in this case:  ${\rm var \,}T \!\rightarrow\! 1/2$ as 
$N \!\rightarrow\! \infty$ for $B \smeq 0$.

{\it Numerical calculations---}
The above predictions are compared, in Figs. 1 to 3, with calculations in
which the Schr\"odinger equation is solved for several billiards using the
methods of Ref. \cite{nummethod}. Typical billiards are sketched in the
figures: reflection symmetric structures are generated from an asymmetric
structure by simply applying the symmetry operators. Note that stoppers
block both the direct transmission and the whispering gallery trajectories
\cite{prl94,epl,Savg}. The theoretical ensemble averages are compared with
numerical energy averages using 100 energies for each $N$ spaced further
than the correlation length. To improve statistics, an additional average
was taken over six similar structures differing in the location of the
stoppers. The classical transmission probability is within $0.005$ of $0.5$
for all structures.

Fig. 1 shows the weak-localization correction (WLC) as a function of $N$ for
$B\smneq 0$. The effect of symmetry is striking: the WLC is zero in the
absence of symmetry, negative for UD symmetry, and positive for LR. 
As noted above, $\langle T \rangle$ is larger than the classical transmission
($N/2$) in the LR case (coherent forward-scattering).

\begin{figure}
\epsfxsize=12cm 
\epsffile[40 251 788 580]{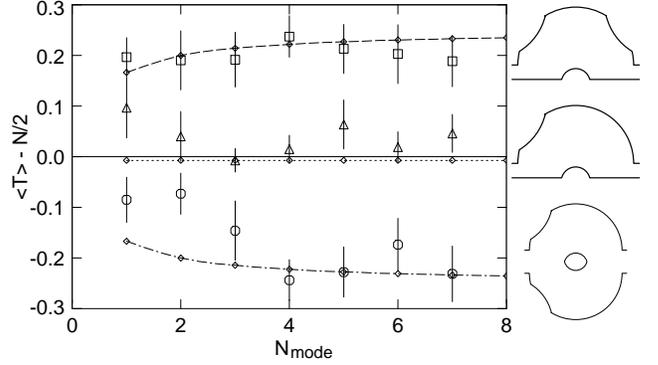}
\caption{
The weak-localization correction for $B \smneq 0$ as a function of the number
of modes in the lead: asymmetric (dotted, triangles), up-down
(dot-dashed, circles), and left-right (dashed, squares) structures.
Lines are theoretical results from Table III (and Ref. \protect\cite{prl94}
for the asymmetric case); 
symbols with
statistical error bars are numerical results averaged over $BA/\phi_0 \smeq 2,4$.
Typical cavities are shown on the side. The presence of symmetry has a strong
effect and can produce either a positive or negative correction.
}
\end{figure}

\begin{figure}[b]
\epsfxsize=12cm 
\epsffile[40 251 788 580]{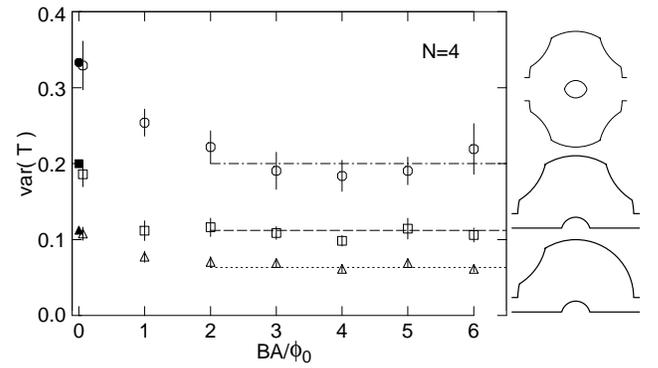}
\caption{Variance of the transmission for $N \smeq 4$ as a function of
magnetic field: asymmetric (dotted, triangles), left-right (dashed,
squares), and four-fold (dot-dashed, circles) structures. Lines and
solid symbols are theoretical results (Table IV and Ref. \protect\cite{prl94}); 
open symbols with statistical error bars are numerical results. 
The magnitude of the
conductance fluctuations increases as the degree of symmetry increases. 
}
\end{figure}

Fig. 2 shows the magnetic field dependence of ${\rm var \,} T$. Applying a
field causes a crossover between symmetry classes. While the present model
does not describe this cross-over, our random matrix theory does predict the
limits $B\smeq 0$ and $BA/\phi_0 \!\gtrsim\! 2$ (as long as 
$\langle S \rangle \!\approx\! 0$--- for much larger fluxes this no longer
holds and we enter a different regime \cite{epl,Savg}). Notice that the
magnitude of the conductance fluctuations increases considerably as one
moves from asymmetric to LR to 4F structures.

The agreement between theory and numerics is very good in both Figs. 1 and
2. In fact, we find that the agreement for ${\rm var \,} T$ is already good
for individual structures. This is good evidence that ${\rm var \,} T$ is a
truely universal quantity for spatially symmetric quantum billiards, as long
as the classical dynamics is strongly chaotic and direct processes are
absent. We cannot say the same for the WLC. While the agreement is good once
we average over six structures, we have found a non-negligible
sample to sample variation. Thus the WLC is of a less
universal character than the variance and may need, for its description,
further sample-specific information.

Fig. 3 shows $w(T)$ for the LR case with $N\smeq 1,2$. These
distributions are very different from those for the asymmetric 
case \cite{prl94,jlp}. There are a number of striking features: 
(a)~for $B\smeq 0$, $N\smeq 1$, the distribution has maxima at 
$T\smeq 0$ and $1$; 
(b)~for $B\smneq 0$, $N\smeq 1$, the maximum is at $T\smeq 1$;
(c)~for $N\smeq 2$, the data near $T\smeq 1$ are consistent with the 
theoretical logarithmic singularity for $B\smeq 0$ and the square-root 
singularity for $B\smneq 0$.

\begin{figure}[b]
\epsfxsize=10cm 
\epsffile[70 250 610 720]{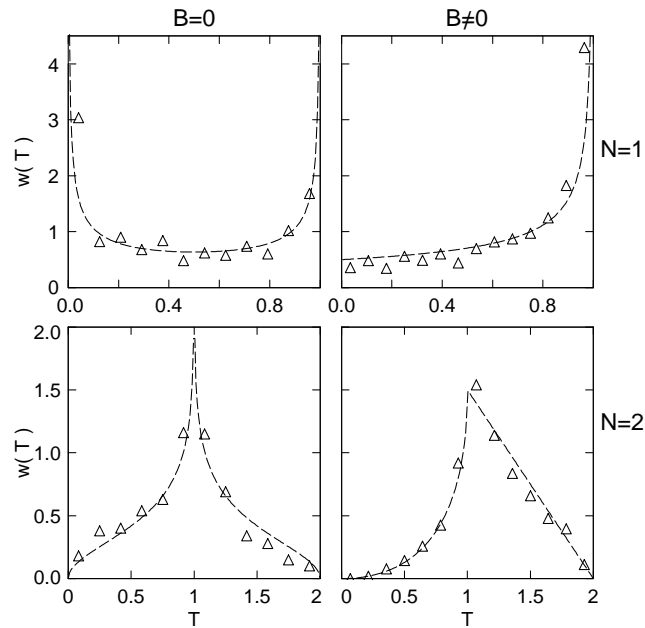}
\caption{The probability density of the transmission for left-right 
symmetry: for $N \smeq 1,2$ and both $B \smeq 0$ and $B \smneq 0$.
The dashed lines are random-matrix theory results from Table V;
the triangles are numerical results.
}
\end{figure}

{\it Summary---}
We have shown, first, that reflection symmetry has a large effect on the
quantum transport properties of classically chaotic billiards and, second,
that the extended circular ensemble random matrix theory describes these
structures well. Because the conductance may couple the different
parity-diagonal blocks of $S$, the quantum transport properties are not a
simple superposition of circular ensemble properties. This work indicates
that the full statistical distribution of the conductance is the same
(``universal''), with some reservations for its centroid, for systems that
(a)~have the same spatial symmetry, (b)~show hard chaos classically, and
(c)~lack direct processes. Experiments observing spatial symmetry effects
should be possible in both microwave cavities and mesoscopic systems. The
effect of the unavoidable deviations from perfect symmetry can be estimated
semiclassically along the lines of Ref. \cite{Ullmo96}, and the results
suggest that the best current material \cite{Marcus,Chang94} is sufficiently
clean.


\vspace*{-0.8cm}
\begin{references}

\bibitem{RevMes}
{\it Mesoscopic Phenomena in Solids}, edited by B.L. Altshuler, P.A. Lee, and
R.A. Webb (North-Holland, New York, 1991).

\bibitem{Marcus}
C.M. Marcus, A.J. Rimberg, R.M. Westervelt, P.F. Hopkins and A.C. Gossard,
Phys. Rev. Lett. {\bf 69}, 506 (1992);
I.H. Chan, {\it et al.},
Phys. Rev. Lett. {\bf 74}, 3876 (1995).

\bibitem{Chang94}
A.M. Chang, H.U. Baranger, L.N. Pfeiffer, and K.W. West, 
Phys. Rev. Lett. {\bf 73}, 2111 (1994).

\bibitem{microwave}
See, e.g., J. Stein, H.-J. St\"ockman, and U. Stoffregen,
Phys. Rev. Lett. {\bf 75}, 53 (1995);
A. Kudrolli, V. Kidambi, and S. Sridhar, 
Phys. Rev. Lett. {\bf 75}, 822 (1995).

\bibitem{uzi}
R. Bl\"umel and U. Smilansky, 
Phys. Rev. Lett. {\bf 60}, 477 (1988); 
{\bf 64}, 241 (1990);
E. Doron and U. Smilansky, Nucl Phys. {\bf A545}, C455 (1992).

\bibitem{hans}
S. Iida, H.A. Weidenm\" uller, and J.A. Zuk, 
Ann. Phys. (N.Y.) {\bf 200}, 219 (1990);
C.H. Lewenkopf and H.A. Weidenm\"uller, 
Ann. Phys. (N.Y.) {\bf 212}, 53 (1991).

\bibitem{prl94}
H.U. Baranger and P.A. Mello, 
Phys. Rev. Lett {\bf 73}, 142 (1994).

\bibitem{jlp}
R.A. Jalabert, J-L. Pichard, and C.W.J. Beenakker, 
Europhys. Lett. {\bf 27}, 255 (1994).

\bibitem{dyson}
F.J. Dyson,
J. Math. Phys. {\bf 3}, 140 (1962); {\bf 3}, 1199 (1962).

\bibitem{mehta-porter}
M.L. Mehta,
{\it Random Matrices} (Academic, New York, 1991);
C.E. Porter, 
{\it Statistical Theories of Spectra: Fluctuations} (Academic, New York, 1965).

\bibitem{hastings}
For the unrealistic but instructive case of reflection-symmetric 
{\it disordered} conductors see
M.B. Hastings, A.D. Stone, and H.U. Baranger,
Phys. Rev. B {\bf 50}, 8230 (1994).

\bibitem{RobBer}
M. Robnik and M.V. Berry,
J. Phys. A {\bf 19}, 669 (1986).

\bibitem{jpa}
V.A. Gopar, M. Mart\'{\i}nez, P.A. Mello, and H.U. Baranger,
J. Phys. A {\bf 29}, 881 (1996).

\bibitem{nummethod}
H.U. Baranger, D.P. DiVincenzo, R.A. Jalabert, and A.D. Stone,
Phys. Rev. B {\bf 44}, 10 367 (1991).

\bibitem{epl}
H.U. Baranger and P.A. Mello, 
Europhys. Lett. {\bf 33}, 465 (1996)

\bibitem{Savg}
Using an average over surface roughness disorder, we find 
$\langle S \rangle \!\approx\! 0$ which justifies our use of the 
invariant measures. The fact that this is true only in structures 
with stoppers highlights the importance of blocking short paths.

\bibitem{Ullmo96}
K. Richter, D. Ullmo, and R.A. Jalabert, 
Phys. Rev. B, in press (1996).

\end{references}
\end{document}